\documentclass[prl,twocolumn,showpacs,superscriptaddress]{revtex4}

\usepackage{graphicx}
\usepackage{dcolumn}

\usepackage{eucal}
\usepackage[dvips]{epsfig}
\usepackage{amssymb}

\usepackage{amsmath,amsthm}

\newcommand{\be}{\begin{eqnarray}}
\newcommand{\ee}{\end{eqnarray}}
\newcommand{\bse}{\begin{subequations}}
\newcommand{\ese}{\end{subequations}}

\newcommand{\bnum}{\begin{enumerate}}
\newcommand{\enum}{\end{enumerate}}

\newcommand{\bit}{\begin{itemize}}
\newcommand{\eit}{\end{itemize}}

\newcommand{\bc}{\begin{cases}}
\newcommand{\ec}{\end{cases}}

\newcommand{\bpm}{\begin{pmatrix}}
\newcommand{\epm}{\end{pmatrix}}

\newcommand{\bvm}{\begin{vmatrix}}
\newcommand{\evm}{\end{vmatrix}}

\newcommand{\bs}{\boldsymbol}

\newcommand{\mcal}{\mathcal}

\newcommand{\mrm}{\mathrm}

\newcommand{\ga}{\alpha}
\newcommand{\gb}{\beta}

\newcommand{\gd}{\delta}

\newcommand{\Gs}{\Sigma}
\newcommand{\Gl}{\Lambda}

\newcommand{\p}{\partial}
\newcommand{\f}{\frac}
\newcommand{\diff}{\mrm{d}}

\newcommand{\lan}{\langle}
\newcommand{\ran}{\rangle}
\newcommand{\bbf}{\mathbb{F}}
\newcommand{\fil}{\mathcal{F}}
\newcommand{\bbg}{\mathbb{G}}
\newcommand{\gcal}{\mathcal{G}}

\newcommand{\csp}{\;,\qquad}

\newcommand{\dP}{\diff P}

\newcommand{\dX}{\diff X}

\newcommand{\dt}{\diff t}
\newcommand{\ds}{\diff s}
\newcommand{\dB}{\diff B}

\newcommand{\dtau}{\diff \tau}

\begin{document}

\title{Time parameters and Lorentz transformations of relativistic stochastic processes}

\author{J\"orn Dunkel}
\email{jorn.dunkel@physics.ox.ac.uk}
\affiliation{Rudolf Peierls Centre for Theoretical Physics, University of Oxford, 1 Keble Road, Oxford  OX1 3TF, U.K.}
\affiliation{Institut
f\"ur Physik, Universit\"at Augsburg, Universit\"atsstra{\ss}e 1,
Augsburg, Germany}
\author{Peter H\"anggi}
\affiliation{Institut
f\"ur Physik, Universit\"at Augsburg, Universit\"atsstra{\ss}e 1,
Augsburg, Germany}
\author{Stefan Weber}
\affiliation{School of Operations Research and Information
Engineering, 279 Rhodes Hall, Cornell University,  Ithaca, NY 14853, USA}

\date{\today}

\begin{abstract}
Rules for the transformation of time parameters in relativistic Langevin equations are derived and discussed. In particular, it is shown that, if a coordinate-time parameterized process approaches the relativistic J\"uttner-Maxwell distribution, the associated proper-time parameterized process converges to a modified  momentum distribution, differing by a factor proportional to the inverse energy.
\end{abstract}

\pacs{
02.50.Ey, 
05.40.-a, 
05.40.Jc, 
47.75.+f  
}

\maketitle


Stochastic processes (SPes) present an ubiquitous tool for modelling complex phenomena  in physics~\cite{1982HaTh,1990HaTaBo,2005FrKr}, biology~\cite{2002Allen,2006Wilkinson}, or  economics and finance~\cite{bachelier,b&s,merton_debt,shreve1}. Stochastic concepts provide a promising alternative to deterministic models whenever the underlying microscopic dynamics of a relevant observable is not known exactly but plausible assumptions about the underlying statistics can be made. A specific area where the formulation of consistent microscopic interaction models becomes difficult~\cite{1929Fo,1945WhFe,1965DaWi} concerns classical relativistic many-particle systems. Accordingly, SPes provide a useful phenomenological approach to describing, e.g., the interaction of a relativistic particle with a fluctuating environment~\cite{1997DeMaRi,2005DuHa,2005Zy,2007AnFr,2008ChDe}. Applications of stochastic concepts to relativistic problems include thermalization processes in quark-gluon plasmas, as produced in relativistic heavy ion colliders~\cite{1988Sv,1997RoEtaAl,2005HeRa,2006RaGrHe}, or complex high-energy processes in astrophysics~\cite{1997MaGo_1,1998ItKoNo,2006WoMe,2006BeLeDe}.
\par
While these applications illustrate the practical relevance of relativistic SPes, there still exist severe conceptual issues which need clarification from a theoretical point of view. Among these is the choice of the time-parameter that quantifies the evolution of a relativistic SP~\cite{1968Ha}.  This problem does not arise within a nonrelativistic  framework, since the Newtonian physics postulates the existence of a universal time which is the same for any inertial observer; thus, it is natural to formulate nonrelativistic SPes by making reference to this universal time. By contrast, in special relativity~\cite{1905Ei_SRT1,SexlUrbantke} the notion of time becomes frame-dependent, and it is necessary to carefully distinguish between different time parameters when constructing relativistic SPes. For example, if the random motion of a relativistic particle is described in a $t$-parameterized form, where $t$ is the time coordinate of some fixed inertial system $\Gs$, then one may wonder if/how this process can be re-expressed in terms of the particle's proper-time $\tau$, and \textit{vice versa}. Another closely related question~\cite{2008ChDe} concerns the problem of how a certain SP appears to a moving observer, i.e.: How does a SP behave under a Lorentz transformation?
\par
The present paper aims at clarifying the above questions for a broad class of relativistic SPes governed by relativistic Langevin equations~\cite{1997DeMaRi,2005DuHa,2005Zy,2007AnFr,2008ChDe}. First, we will discuss a heuristic approach that suffices for most practical calculations and clarifies the basic ideas. Subsequently, these heuristic arguments will be substantiated with a mathematically rigorous foundation by applying theorems for the time-change of (local) martingale processes~\cite{protter}. The main results can be summarized as follows:
If a relativistic Langevin-It{\^o} process has been specified in the inertial frame $\Gs$ and is parameterized by the associated $\Gs$-coordinate time $t$, then this process can be reparameterized by  its proper-time $\tau$ and the resulting process is again of the Langevin-It{\^o} type. Furthermore, the process can be Lorentz transformed to a moving frame $\Gs'$, yielding a Langevin-It{\^o} process that is parameterized by the $\Gs'$-time~$t'$. In other words, similar to the case of purely deterministic relativistic equations of motions, one can choose freely between different time parameterizations in order to characterize these relativistic SPes -- but the noise part needs to be transformed differently than the deterministic part.

\paragraph{Notation.--}
We adopt the metric convention $(\eta_{\ga\gb})=\mrm{diag}(-1,1,\ldots,1)$ and units such that the speed of light $c=1$. Contra-variant space-time and energy-momentum four-vectors are denoted by $(x^\ga)=(x^0,x^i)=(x^0,\bs x)=(t,\bs x)$ 
and $(p^\ga)=(p^0,p^i)=(p^0,\bs p)$, respectively, with Greek indices $\ga=0,1,\ldots,d$ and Latin indices $i=1,\ldots, d$, where $d$ is the number of space dimensions. 
Einstein's summation convention is applied throughout.

\paragraph{Relativistic Langevin equations.--}
As a starting point, we consider the $t$-parameterized random motion of a relativistic particle (rest mass $M$) in the inertial lab frame $\Gs$. The lab frame is defined by the property that the thermalized background medium (heat bath) causing the stochastic motion of the particle is  at rest in $\Gs$ (on average). We assume that the particle's trajectory $(\bs X(t),\bs P(t))=(X^i(t),P^i(t))$ in $\Gs$ is governed by a stochastic differential equation (SDE) of the form~\cite{1997DeMaRi,2005DuHa,2005Zy,2007AnFr,2008ChDe}
\bse\label{e:RLE}
\be\label{e:RLE-a}
\dX^\ga(t)
&=&(P^\ga/{P^0})\; \dt,\\
\dP^i(t)\label{e:RLE-b}
&=& {A^i}\; \dt + {C^i}_j  \diff B^j(t).
\ee
\ese
Here, $\dX^0(t)=\dt$ and $\dX^i(t):=X^i(t+\diff t)-X^i(t)$ denote the time and  position increments, $\dP^i(t):=P^i(t+\diff t)-P^i(t)$ the momentum change. $P^0(t):=(M^2+ \bs P^2)^{1/2}$ is the relativistic energy, and  $V^i(t):=\diff X^i/\diff t=P^i/P^0$ are the velocity components in $\Gs$. In general, the functions $A^i$ and ${C^i}_j$ may depend on the time, position and momentum coordinates of the particle. The random driving process $\bs B(t)=(B^j(t))$ is taken to be a $d$-dimensional $t$-parameterized standard Wiener process~\cite{1923Wi, KaSh91, protter}, i.e.,  $\bs B(t)$ has continuous paths, for $s>t$ the increments are normally distributed,
\be\label{e:OUP_langevin_math_density}
\mcal{P}\{\bs B(s)-\bs B(t) \in [\bs u, \bs u+\diff\bs u]\}=
\f{ e^{ -{|\bs u|^2}/{[2\,(s-t)]} }  }
{[2\pi\,(s-t)]^{d/2}}
\;\diff^d u,
\ee
and independent for non-overlapping time intervals \footnote{For simplicity, we have assumed that $\bs B (t)$  is $d$-dimensional, implying  that $C^i_{~j}$ is a square matrix. However, all results still hold if $\bs B(t)$ has a different dimension.}. 
\par
Upon naively dividing Eq.~\eqref{e:RLE-b} by $\dt$, we see that $A^i$ can be interpreted as a deterministic force component, while ${C^i}_j  \diff B^j(t)/\dt$ represents random \lq noise\lq. However, for the Wiener process  the derivatives $\diff B^j(t)/\dt$ are not well-defined mathematically so the differential representation~\eqref{e:RLE} is in fact short hand for a stochastic integral equation~\cite{KaSh91, protter} with ${C^i}_j \dB^j$ signifying an infinitesimal increment of the It{\^o} integral~\cite{1944Ito,1951Ito}. Like a deterministic integral, stochastic integrals can be approximated by Riemann-Stieltjes sums but the coefficient functions need to be evaluated at the \emph{left} end point $t$ of any time interval $[t, t+\dt]$ in the It{\^o} discretization~\footnote{One could also consider other discretization rules~\cite{1965Fisk,1966St,1982HaTh,KaSh91, protter}, but then the rules of stochastic differential calculus must be adapted.}.
In contrast to other discretization rules~\cite{1965Fisk,1966St,1982HaTh,KaSh91,protter}, the It{\^o} discretization implies that the mean value of the noise vanishes, i.e., $ \lan{C^i}_j  \diff B^j(t)\ran=0$ with $\lan\,\cdot\; \ran$ indicating an average over all realizations of the Wiener process $\bs B(t)$. In other words, It{\^o} integrals with respect to $\bs B(t)$ are (local) martingales~\cite{protter}.
Upon applying It{\^o}'s formula~\cite{KaSh91, protter} to the mass-shell condition $P^0(t)=(M^2+\bs P^2)^{1/2}$, one can derive from  Eq.~\eqref{e:RLE-b} the following equation for the relativistic energy:
\be
&&\dP^0(t)
=
A^0\,\dt +
{C^0}_r\diff B^r(t)
,\label{e:RLE_0-b}
\\\notag
&&A^0:= \frac {A_i P^i}{P^0} + \f{D_{ij}}{2}
\left[  \f{\delta^{ij}}{P^0} - \f{P^i P^j}{(P^0)^3} \right],\quad
C^0_{~j}  :=  \f{P^i C_{ij}}{P^0}, 
\ee
where $A_i:=A^i$, $D_{ij}:=D^{ij}=\sum_r C^i_r C^j_r$ and $C_{ir}:={C^i}_r$.
\par
Equations~\eqref{e:RLE} define a  straightforward relativistic  generalization~\cite{1997DeMaRi,2005DuHa,2005Zy} of the classical Ornstein-Uhlenbeck process~\cite{1930UhOr}, representing a standard model of Brownian motion theory~\footnote{In the nonrelativistic limit $c\to\infty$, $P^0\to  M$ in Eq.~\eqref{e:RLE-a}.}. The structure of Eq.~\eqref{e:RLE-a} ensures that the velocity remains bounded, $|\bs V|<1$, even if the momentum $\bs P$ were to become infinitely large. 
When studying SDEs of the type~\eqref{e:RLE}, one is typically interested in the probability $f(t,\bs x,\bs p)\,\diff^d x \diff^d p$ of finding the particle at time $t$ in the infinitesimal phase space interval $[\bs x,\bs x+\diff \bs x]\times [\bs p,\bs p+\diff \bs p]$. Given Eqs.~\eqref{e:RLE}, the non-negative, normalized probability density $f(t,\bs x,\bs p)$ is governed by the Fokker-Planck equation (FPE)
\be\label{e:FPE}
\biggl(\f{\p}{\p t} +\f{p^i}{p^0}\f{\p}{\p x^i}\biggr) f
=
\f{\p}{\p p^i}\biggl[-A^i f +\f{1}{2}
\f{\p}{\p p^k} \bigl(D^{ik} f \bigr)\biggr],
\ee
where $f$ is a Lorentz scalar~\cite{1969VK} and $p^0=(M^2+\bs p^2)^{1/2}$  \footnote{Equation~\eqref{e:FPE} is not covariant, because we are considering here 
the \lq true\rq~phase space density $f(t,\bs x,\bs p)$ rather than the \lq extended\rq~phase space density $\tilde f(t,\bs x, p^0,\bs p)$.}. Deterministic initial data $\bs X(0)=\bs x_0$ and $\bs P(0)=\bs p_0$ translates into the localized initial condition $f(0,\bs x,\bs p)=\gd(\bs x-\bs x_0)\;\gd(\bs p-\bs p_0)$.
Physical constraints on the coefficients $A^i(t,\bs x,\bs p)$ and ${C^i}_r(t,\bs x,\bs p)$ may arise from symmetries and/or thermostatistical considerations. For example, neglecting additional external force fields and considering a heat bath that is stationary, isotropic and position independent in $\Gs$, one is led to the ansatz
\bse\label{e:constraints}
\be
A^i=-\ga(p^0)\,p^i \csp
{C^i}_j=[2D(p^0)]^{1/2}\;{\gd^i}_j.
\ee
where the friction and noise coefficients $\ga$ and $D$ depend on the energy $p^0$ only. Moreover, if the stationary momentum distribution is expected to be a thermal J\"uttner function~\cite{1911Ju,2007CuEtAl}, i.e., if $f_\infty:=\lim_{t\to\infty } f \propto \exp(-\gb p^0)$ in $\Gs$, then $\ga$ and $D$ must satisfy the fluctuation-dissipation condition~\cite{1997DeMaRi}
\be\label{e:einstein}
0\equiv
\ga(p^0)\, p^0 + \diff D(p^0)/\diff p^0 -\gb D(p^0).
\ee
\ese
In this case, one still has the freedom to adapt one of the two functions $\ga$ or $D$.
\par
In the remainder, we shall discuss how the process~\eqref{e:RLE} can be reparameterized in terms of
its proper-time~$\tau$, and how it transforms under the proper Lorentz group~\cite{SexlUrbantke}.

\paragraph{Proper-time parameterization.--}
The stochastic proper-time differential $\dtau(t)=(1-\bs V^2)^{1/2}\dt$ may be expressed as
\bse\label{e:proper-time-heuristics}
\be\label{e:proper-time-heuristics-a}
\dtau(t)=(M/P^0)\,\dt.
\ee
The inverse of the function $\tau$ is denoted by $\hat X^0(\tau) = t(\tau)$ and represents  the time coordinate of the particle in the inertial frame $\Sigma$, parameterized by the proper time~$\tau$. Our goal is to find SDEs for the reparameterized processes $\hat{X}^\ga(\tau):= X^\ga(t(\tau))$ and $\hat{P}^\ga(\tau)= P^\ga(t (\tau))$ in~$\Gs$. The  heuristic derivation is based on the relation
\be
\dB^j(t)
\simeq
\sqrt{\dt}
=
\biggr(\f{\hat{P}^0}{M}\biggr)^{1/2}\sqrt{\dtau}
\simeq
\biggr(\f{\hat{P}^0}{M}\biggr)^{1/2}\diff\hat{B}^j(\tau),
\quad
\label{e:proper-time-heuristics-b}
\ee
\ese
where $\hat{B}^j(\tau)$ is a standard Wiener process with time-parameter $\tau$. The rigorous justification of Eq.~\eqref{e:proper-time-heuristics-b} is given below. 
Inserting Eqs.~\eqref{e:proper-time-heuristics} in Eqs.~\eqref{e:RLE} one finds
\bse\label{e:RLE-tau}
\be
\diff \hat X^\ga(\tau)&=&(\hat P^\ga/M)\,\dtau,\\
\diff\hat P^i(\tau) &=&
\hat A^i\; \dtau + {\hat C^i}_{~j} \diff \hat B^j(\tau),
\ee
\ese
where $\hat A^i : =   (\hat P^0/M)\, A^i(\hat X^0, \hat{\bs X}, \hat {\bs  P})$ and $\hat C^i_{~j} := (\hat P^0 / M)^{1/2}\; C^i_{~j}(\hat X^0, \hat{\bs X}, \hat{\bs P})$.
The  FPE for  the associated probability density $\hat{f}(\tau, x^0, \bs x, \bs p)$ reads
\be\label{e:FPE-tau}
\biggl(\f{\p}{\p \tau} +\f{p^\ga}{M}\f{\p}{\p x^\ga}\biggr) \hat{f}
=
\f{\p}{\p p^i}\biggl[-\hat{A}^i \hat{f} +\f{1}{2}
\f{\p}{\p p^k} \bigl({{\hat D}^{ik}}\hat{f}\bigr)\biggr]
\ee
where now ${\hat D}^{ik}:=\sum_r{\hat{C}^i}_r {\hat{C}^k}_r$. We note that $\hat{f}(\tau,x^0,\bs x,\bs p)\;\diff x^0\diff^d x \diff^d p$ gives probability of finding the particle at proper-time~$\tau$ in the interval $[t,t+\dt]\times [\bs x,\bs x+\diff \bs x]\times [\bs p,\bs p+\diff \bs p]$ in the inertial frame $\Sigma$. 

\par
\begin{figure}[t]
\centering
\epsfig{file=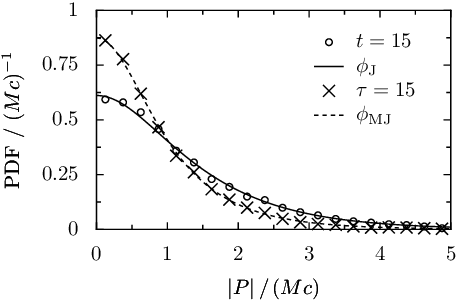, width=7.5cm}
\caption{\label{fig01}
\lq Stationary\rq\space probability density function (PDF) of the absolute momentum $|P|$ measured at time $t=15$ ($\times$) and $\tau=15$ ($\circ$) from $10000$ sample trajectories of the one-dimensional $(d=1)$ relativistic Ornstein-Uhlenbeck process~\cite{1997DeMaRi}, corresponding to coefficients $D(p^0)=const$ and $\ga(p^0)=\gb D/p^0 $ in Eqs.~\eqref{e:RLE} and \eqref{e:constraints}. Simulation parameters: $\dt=0.001$, $M=c=\gb=D=1$.}
\end{figure}

Remarkably, if the coefficient functions satisfy the constraints~\eqref{e:constraints} --  so that the stationary solution $f_\infty$ of Eq.~\eqref{e:FPE} is a J\"uttner function  $\phi_\mrm{J}(\bs p)=\mcal{Z}^{-1}\exp(-\gb  p^0)$ -- then the stationary solution $\hat{f}_\infty$ of the  corresponding proper-time FPE~\eqref{e:FPE-tau} is given by a modified J\"uttner function $\phi_{\mrm{MJ}}(\bs p)=\hat{\mcal{Z}}^{-1}\exp(-\gb p^0)/p^0$. The latter can be derived from a relative entropy principle, using a Lorentz invariant reference measure in momentum space~\cite{2007DuTaHa_2}. Physically, the difference between $f_\infty$ and $\hat{f}_\infty$ is due to the fact that measurements at $t=const$ and $\tau=const$ are non-equivalent even if $\tau,t\to \infty.$ This can also be confirmed by direct numerical simulation of Eqs.~\eqref{e:RLE}, see Fig.~\ref{fig01}.
\par
Having discussed the proper-time reparameterization, we next show that a similar reasoning can be applied to transform the SDEs~\eqref{e:RLE} to a moving frame $\Gs'$~\cite{2008ChDe}.
\paragraph{Lorentz transformations.--}
Neglecting time-reversals, we consider a proper Lorentz transformation~\cite{SexlUrbantke} from the lab frame $\Gs$ to $\Gs'$, mediated by a constant matrix ${\Gl^\nu}_\mu$ with ${\Gl^0}_0>0$, that leaves the metric tensor $\eta_{\ga\gb}$ invariant.  We proceed in two steps: First we define $$
Y'^\nu(t):={\Gl^\nu}_\mu  X^\mu(t)\csp
G'^\nu(t):={\Gl^\nu}_\mu  P^\mu(t).
$$
Then we replace $t$ by the coordinate time $t'$ of $\Gs'$ to obtain processes $X'^\ga(t')= Y'^\ga(t(t'))$ and $P'^\ga(t')= G'^\ga(t(t'))$. Note that $\dt'(t)
=\diff Y'^0(t)={\Gl^0}_\mu  \dX^\mu(t)$, and, hence,
\be \label{e:LT-heuristics-1}
\dt'(t)
= \f{{\Gl^0}_\mu P^\mu}{P^0}\,\dt
=\f{G'^0}{P^0}\,\dt
= \frac{P'^0(t'(t))}{(\Lambda^{-1})^0_{\;\mu} P'^\mu(t'(t))}\, \dt,
\ee
where $\Gl^{-1}$ is the inverse Lorentz transformation. Thus, a similar heuristics as in Eq.~\eqref{e:proper-time-heuristics-b} gives
$$
\dB^j(t)
\simeq\notag
\sqrt{\dt}
=
\biggl(\f{P^0}{P'^0}\biggr)^{\!\!\!1/2}
\!\!\!\!\!\sqrt{\dt'}
\simeq
\biggl[\f{(\Gl^{-1})^0_{\;\mu} P'^\mu}{P'^0}\biggr]^{\!1/2}
\!\!\!\!\!
\diff B'^j(t'),
$$
where $B'^j(t')$ is a Wiener process with time~$t'$. Furthermore, defining primed 
coefficient functions in $\Sigma'$ by
\bse
\be
&&\notag
 A'^i(x'^0, {\bs x'}, {\bs p'})  := [(\Lambda^{-1})^0_{~\mu} p'^\mu/{p'^0}] \;
\times\\
&&\quad\notag
\Lambda^i_{~\nu}\, A^\nu \left( (\Lambda^{-1})^0_{~\mu} x'^\mu,  (\Lambda^{-1})^i_{~\mu} x'^\mu, (\Lambda^{-1})^i_{~\mu} p'^\mu  \right),
\qquad
\\
&& \notag
C'^i_{~j}(x'^0, {\bs x'}, {\bs p'}) :=
[(\Lambda^{-1})^0_{~\mu} p'^\mu/{p'^0}]^{1/2}
\times
\\&&
\quad\notag
\Lambda^i_{~\nu} \, C^\nu_{~j} \left( (\Lambda^{-1})^0_{~\mu} x'^\mu,  (\Lambda^{-1})^i_{~\mu} x'^\mu, (\Lambda^{-1})^i_{~\mu} p'^\mu  \right),
\ee
\ese
the particle's trajectory $(\bs X'(t'),\bs P'(t'))$ in $\Gs'$ is again governed by a SDE of the standard form
\bse\label{e:RLE'}
\be\label{e:RLE'-a}
\dX'^\ga(t')
&=&(P'^\ga/{P'^0})\; \dt',\\
\dP'^i(t')\label{e:RLE'-b}
&=& {A'^i}\; \dt' + {C'^i}_j \, \diff B'^j(t').
\ee
\ese

\paragraph{Rigorous justification.--}
We will now rigorously derive the transformations of SDEs under time changes and thereby show that the heuristic transformations leading to Eqs.~\eqref{e:RLE-tau} and ~\eqref{e:RLE'} are justified; i.e., we are interested in a time-change $t\mapsto \breve{t}$ of a generic SDE
\bse\label{e:trafo}
\be\label{e:sde}
\diff Y(t) = E\,\dt + F_j\, \dB^j(t),
\ee
where $E$ and $F_j$ will typically be smooth functions of the state-variables ($Y,\ldots$)~\footnote{The state variables of the system are assumed to have continuous paths and need to satisfy suitable integrability conditions. More generally, $E=E(t)$ and $F_j= F_j(t)$ can be assumed to be continuous adapted processes.}, and ${\bs B} (t)=(B^j(t))$ is a $d$-dimensional standard Wiener process~\footnote{The Wiener process is defined on a complete filtered probability space $(\Omega, \fil, \bbf, P)$ that satisfies the usual hypotheses~\cite{protter}. The increasing family $\bbf=(\fil_t)$ is called a filtration. $\fil_t$ denotes the information that will be available to an observer at time $t$ who follows the particle.}. We consider a time-change  $t\mapsto \breve{t}$ specified in the form [cf. Eqs.~\eqref{e:proper-time-heuristics-a}~and~\eqref{e:LT-heuristics-1}]
\be \label{e:t-trafo}
\diff\breve{t} = H\; \dt,
\qquad
\breve{t}(0) =  0,
\ee 
with $H$ being a strictly positive smooth function~\footnote{More precisely, in general $H=H(t)$ is a strictly positive, continuous adapted process such that $\mcal{P}[ \int_0^t H(s) \diff s < \infty \forall t] =1$  and $\mcal{P}[  \int _0^\infty H(s) \diff s =\infty ] =1$.} of~$(Y,\ldots)$. The inverse of  $\breve{t}(t)$ is denoted by $t(\breve{t})$. We would like to show that Eq.~\eqref{e:sde} can be rewritten as
\be\label{e:sde-goal}
\diff\breve{Y}(\breve{t}) = \breve{E}\,\diff\breve{t} + \breve{F}_j\, \diff\breve{B}^j(\breve{t}),
\ee
where $\breve{Y}(\breve{t}) :=  Y (t(\breve{t}))$, $\breve{E}(\breve{t}) :=  {E (t(\breve{t}) )}/{H (t(\breve{t}) )}$,
\linebreak 
$\breve{F}^j(\breve{t}) :=  {F^j (t(\breve{t}) )}/{\sqrt{H (t(\breve{t}) )}}$, and 
\be\label{e:b-trafo}
\diff\breve{B}^j (\breve{t}) := \sqrt{H}\; \dB^j (t)
\ee
\ese
is a $d$-dimensional Wiener process with respect to the new time parameter~$\breve{t}$~\footnote{The information available to an observer of the particle at time $\breve{t}$ is denoted by $\gcal_{\breve{t}}$. The corresponding filtration is denoted by $\bbg=(\gcal_{\breve{t}})$; cf. Chapt. I.1 in \cite{protter}. The mathematically precise statement regarding the time-change is that $\bs{\breve{B}}(\breve{t})$ is a standard Wiener process with respect~to~$\bbg$.}.
\par
First, we need to prove that Eq.~\eqref{e:b-trafo} or, equivalently, 
 $\breve{B}^j(\breve{t}) := \int_0^{t(\breve{t})}\; \sqrt{H(s)}\; \diff B^j(s)$ does indeed define a Wiener process. To this end, we note that for fixed $j\in\{1, \dots, d\}$ the process $L^j(t) := \int_0^t \sqrt{H(s)}\, \dB^j(s)$ is a continuous local martingale, whose quadratic variation
$$
[L^j, L ^j](t):= 
\lim_{n\to\infty} \sum_{k=0}^{2^n-1} \left\{  L^j\left(
\frac{(k+1) t} {2^n}  \right)  - L^j\left( \frac {k t} {2^n}  \right) 
\right\}^2
$$
is given by $[L^j, L^j](t)=\int_0^t H(s) \diff s$~\footnote{Convergence is uniform on compacts in probability; see \cite{protter} for a definition of the quadratic covariation~$[L^i,L^j]_t$.}. For the quadratic variation of  $\breve{B}^j(\breve{t}) = L^j(t(\breve{t}))$ we then obtain $[\breve{B}^j, \breve{B}^j](\breve{t}) = [L^j, L^j](t(\breve{t})) = \int_0^{t(\breve{t})} H(s)\, \ds = \breve{t}$. For $i\neq j$, we have $
[\breve{B}^j, \breve{B}^i](\breve{t}) = \int_0^{t(\breve{t})} H(s)\, \diff[B^j, B^i](s) = 0.
$
Thus, L{\'e}vy's Theorem~\footnote{See Theorem II.8.40, p. 87 in Protter~\cite{protter}.} implies that $\bs{\breve{B}} (\breve{t})= (\breve{B}^j(\breve{t}))$ is a $d$-dimensional standard Wiener process.

Finally, using the definitions of $\breve{Y}$, $\breve{E}$, and $\breve{F}^j$, we find~\footnote{The second equality follows from  Eqs.~\eqref{e:t-trafo} and~\eqref{e:b-trafo} by approximating the processes $E$ and $F^j$ by simple predictable processes, see p. 51 and Theorem II.5.21 in~\cite{protter}.}
\be
\breve{Y}(\breve{t})
&=& \notag
\int_0^{t(\breve{t})} E(s)\, \ds + \int_0^{t(\breve{t})} F_j(s)\,\dB^j(s)
\\
&=& \notag
\int _0^{\breve{t}} \frac{E(t({\breve{s}}))}{H(t(\breve{s}))} d\breve{s} +
\int _0^{\breve{t}}   \frac{F_j(t({\breve{s}}))}{\sqrt{H(t(\breve{s}))}}\, \diff\breve{B}^j(\breve{s})
\\
&= &
\int _0^{\breve{t}} \breve{E}(\breve{s})\, \diff\breve{s} +
\int _0^{\breve{t}} \breve{F}_j({\breve{s}})\,\diff\breve{B}^j(\breve{s}),
\ee
which is just the SDE~\eqref{e:sde-goal} written in integral notation.

\paragraph{Conclusions.--}
The above discussion shows how relativistic Langevin equations can be Lorentz transformed and reparameterized within a common framework. Thus, mathematically, the special relativistic Langevin theory~\cite{1997DeMaRi,2005DuHa,2005Zy,2007AnFr,2008ChDe} is now as  complete as the classical theories of nonrelativistic Brownian motions and deterministic relativistic motions, respectively, both of which are included as special limit cases. From a physics point of view, the most remarkable  observation consists in the fact that the $\tau$-parameterized Brownian motion converges to a modified J\"uttner function~\cite{2007DuTaHa_2} if the corresponding $t$-parameterized process converges to a J\"uttner function~\cite{1911Ju}.  This illustrates that it is necessary to distinguish different notions of \lq stationarity\rq~in special relativity. While the $t$-parameterization appears more natural when describing diffusion processes from the viewpoint of an external observer~\cite{1988Sv,1997RoEtaAl,2005HeRa,2006RaGrHe,1997MaGo_1,1998ItKoNo,2006WoMe,2006BeLeDe}, the $\tau$-parameterization is more convenient when extending the above theory to include particle creation/annihilation  processes, because a particle's lifetime is typically  quantified in terms of its proper-time~$\tau$.  Last but not least, the proper-time parameterization paves  the way toward generalizing the above concepts to general relativity.


\end{document}